\documentclass[floats,floatfix,amssymb,prd,superscriptaddress,nofootinbib,twocolumn,aps]{revtex4-1}
\usepackage{graphicx, epsfig, amssymb} 
\usepackage{amsmath, amsfonts}
\usepackage{bm} 

\usepackage[linktocpage]{hyperref}
\usepackage[caption=false]{subfig}
\usepackage[usenames]{color}

\def\be{\begin{equation}}
\def\ee{\end{equation}}

\newcommand{\bea}{\begin{eqnarray}}
\newcommand{\eea}{\end{eqnarray}}
\newcommand{\ben}{\begin{enumerate}}
\newcommand{\een}{\end{enumerate}}
\newcommand{\bi}{\begin{itemize}}
\newcommand{\ei}{\end{itemize}}

\newcommand{\nn}{\nonumber}

\def\ga{\mathrel{\raise.3ex\hbox{$>$\kern-.75em\lower1ex\hbox{$\sim$}}}}
\def\la{\mathrel{\raise.3ex\hbox{$<$\kern-.75em\lower1ex\hbox{$\sim$}}}}

\def\I_M{{I_{\scriptscriptstyle M\times M}}}

\def\be{\begin{equation}}
\def\ee{\end{equation}}
\def\bea{\begin{eqnarray}}
\def\eea{\end{eqnarray}}
\newcommand{\beq}{\begin{eqnarray}}
\newcommand{\eeq}{\end{eqnarray}}

\newcommand{\beqal}{\begin{eqnarray}\label}
\newcommand{\beqa}{\begin{eqnarray}}
\newcommand{\eeqa}{\end{eqnarray}}

\newcommand {\f}{\frac}

\begin{document}

\title{\large Floating and sinking: the imprint of massive scalars around rotating black holes}
 
\author{Vitor Cardoso}
\affiliation{CENTRA, Departamento de F\'{\i}sica, Instituto Superior T\'ecnico, Universidade T\'ecnica de Lisboa - UTL,
Av.~Rovisco Pais 1, 1049 Lisboa, Portugal.}
\affiliation{Department of Physics and Astronomy, The University of Mississippi, University, MS 38677, USA.}

\author{Sayan Chakrabarti}
\affiliation{CENTRA, Departamento de F\'{\i}sica, 
Instituto Superior T\'ecnico, Universidade T\'ecnica de Lisboa - UTL,
Av.~Rovisco Pais 1, 1049 Lisboa, Portugal.}

\author{Paolo Pani}
\affiliation{CENTRA, Departamento de F\'{\i}sica, Instituto Superior T\'ecnico, Universidade T\'ecnica de Lisboa - UTL,
Av.~Rovisco Pais 1, 1049 Lisboa, Portugal.}

\author{Emanuele Berti}
\affiliation{Department of Physics and Astronomy, The University of Mississippi, University, MS 38677, USA.}
\affiliation{California Institute of Technology, Pasadena, CA 91109, USA}

\author{Leonardo Gualtieri}
\affiliation{Dipartimento di Fisica, Universit\`a di Roma ``Sapienza'' \& Sezione, INFN Roma1, P.A. Moro 5, 00185, Roma, Italy.}


\begin{abstract}
We study the coupling of massive scalar fields to matter in orbit around
rotating black holes. It is generally expected that orbiting bodies will lose
energy in gravitational waves, slowly inspiralling into the black
hole. Instead, we show that the coupling of the field to matter leads to a
surprising effect: because of superradiance, matter can hover into ``floating
orbits'' for which the net gravitational energy loss at infinity is entirely
provided by the black hole's rotational energy. Orbiting bodies remain
floating until they extract sufficient angular momentum from the black hole,
or until perturbations or nonlinear effects disrupt the orbit. For slowly
rotating and nonrotating black holes floating orbits are unlikely to exist,
but resonances at orbital frequencies corresponding to quasibound states of
the scalar field can speed up the inspiral, so that the orbiting body
``sinks''. These effects could be a smoking gun of deviations from general
relativity.
\end{abstract}


\maketitle
\date{today}
\noindent{\bf{\em I. Introduction.}}
Massive scalars are ubiquitous in physics. For example, light scalars spanning several orders of magnitude in mass are predicted in string-theory scenarios \cite{Arvanitaki:2009fg,Arvanitaki:2010sy,Kodama:2011zc}. Massive scalars are observationally viable in scalar-tensor generalizations of Einstein's general relativity \cite{Wagoner:1970vr} and can be regarded as an effective propagating degree of freedom in $f(R)$ theories \cite{Hersh:1985hz,DeFelice:2010aj}. 
In this paper we consider generic massive scalar fields coupled to matter in orbit around a rotating black hole (BH). 

A well-known phenomenon in BH physics is the Penrose process (for particles) and the associated superradiant amplification (for waves) \cite{zeldovich,Bekenstein:1973mi}. Consider a Kerr BH of mass $M$, angular momentum $J=aM$ and horizon radius $r_+=M+\sqrt{M^2-a^2}$, so that the angular velocity of the horizon $\Omega_H=a/2Mr_+$ (here and below we set $G=c=1$). A wave with frequency $\omega<m\Omega_H$ incident on the BH (where $m$ is the azimuthal quantum number) is amplified in a scattering process, the excess energy coming from the BH's rotational energy. 
%
\begin{figure}[bh]
\begin{center}
\epsfig{file=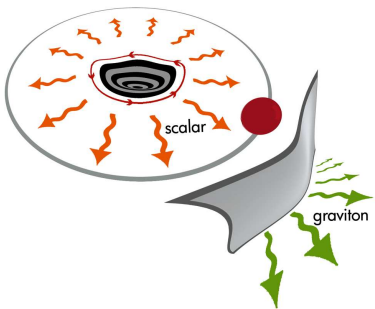,width=6cm,angle=0,clip=true}
\caption{Pictorial description of floating orbits. An orbiting body excites superradiant scalar modes close to the BH horizon. Since the scalar field is massive, the flux at infinity consists solely of gravitational radiation. 
\label{fig:superradiance}}
\end{center}
\end{figure}
Superradiance is responsible for many interesting effects~\cite{Misner:1972kx,Press:1972zz,Damour:1976kh,Cardoso:2004nk,Cardoso:2005vk,Cardoso:2004hs,Friedman:1978,Cardoso:2007az}.
Here we explore the interesting possibility that an object in orbit around a rotating BH may excite superradiant modes to appreciable amplitudes.
As the object orbits around the BH it loses energy in gravitational waves, slowly spiralling in, as shown experimentally by the Hulse-Taylor binary pulsar. This follows from energy balance: if the orbital energy of the particle is $E_p$, and the total (gravitational plus scalar) energy flux is $\dot E_T= \dot E^{g}+\dot E^{s}$, then
\begin{equation}
\dot E_p+\dot E^{g}+\dot E^{s}=0\,.\label{balance}
\end{equation}
Usually $\dot E^{g}+\dot E^{s}>0$, and therefore the orbit shrinks with time. However it is possible that, due to superradiance, $\dot E^{g}+\dot E^{s}=0$. In this case $\dot E_p=0$, and the particle can hover in a ``floating orbit'' \cite{Misner:1972kx,Press:1972zz}.
Here we show that floating orbits, for which the net gravitational energy loss
at infinity is entirely provided by the BH's rotational energy, can exist for a wide range of scalar-field masses. Orbiting bodies will float until they extract sufficient angular momentum from the BH or until disruptive (perhaps nonlinear) effects stop the process. When the BH rotates slowly the condition for superradiance at these resonances is not met, but we show that resonances at small orbital frequencies (corresponding to large positive scalar fluxes going into the horizon) still exist, and that they cause the object to inspiral faster.

\noindent{\bf{\em IIA. Setup.}}
%
The process we consider is quite general. It occurs in all theories of gravity with Kerr BHs as background solutions and a scalar field of mass $\hbar\mu_s$ coupled to matter (see e.g.~\cite{Thorne:1971,Psaltis:2007cw}). 
At first order in perturbation theory, the field equations for the scalar field reduce to 
\be
\left[\square-\mu_s^2\right]\varphi=\alpha {\cal T}\label{EQBD2b}\,.
\ee
Our main results will be to a large extent independent of the source term on the right-hand side, but for concreteness we focus on source terms of the form
\be
{\cal T}=\int \frac{d\bar\tau}{\sqrt{-\bar g^{(0)}}}\,m_p\delta^{(4)}\left(x-X(\bar\tau)\right)\,,
\ee
corresponding to the trace of the stress-energy tensor of a point particle with mass $m_p$, where $\bar g^{(0)}$ is the background (Kerr) metric. In scalar-tensor theories, for example, $\alpha=\sqrt{{8 \pi}/(2+\omega_{\text{BD}})} \left(s-1/2\right)$, where $\omega_{\rm BD}$ is the Brans-Dicke (BD) parameter\footnote{Measurements of the Shapiro time delay require $\omega_{\rm BD}> 40,000$ for $\mu_s=0$ \cite{Will:2005va}, but couplings of order $\omega_{\rm BD}\sim {\cal O}(1)$ are observationally allowed when $\mu_s\gtrsim 10^{-17}$~eV, and no bounds on $\omega_{\rm BD}$ exist when $\mu_s\gtrsim 10^{-16}$~eV~\cite{Perivolaropoulos:2009ak}. Considering a supermassive BH of mass $M\sim 10^5 M_\odot$ and a typical sensitivity $s\sim0.2$, these bounds translate into $\alpha\lesssim 8\cdot 10^{-3}$ when $\mu_s=0$, $\alpha\lesssim 0.9$ when $\mu_s M>10^{-2}$ and no bounds on $\alpha$ when $\mu_s M>0.1$.} and $s$ is an object-dependent ``sensitivity factor''~\cite{Wagoner:1970vr,Ohashi:1996uz}.

Weak-field gravitational radiation circularizes the orbit (see below for a proof in the present context), so we consider equatorial circular orbits around a Kerr BH, but most results apply to more general orbits. 
Using the ``adiabatic approximation'' we assume that the radiation reaction timescale is much longer than the orbital timescale, and compute the total energy flux $\dot E_T$ for geodesic orbits. For prograde orbits, energy, angular momentum and frequency of a particle at $r=r_0$ read
\begin{eqnarray}
 E_p&=&\frac{a \sqrt{M}+\sqrt{r_0} (r_0-2M)}{r_0^{3/4} \sqrt{2 a \sqrt{M}+\sqrt{r_0} (r_0-3M)}}\,m_p\,,\label{Ep}\\
L_p&=&\frac{\sqrt{M} \left(r_0^2-2 a \sqrt{M r_0}+a^2\right)}{r_0^{3/4} \sqrt{2 a \sqrt{M}+\sqrt{r_0} (r_0-3M)}}\,m_p \,,\label{Lp}\\
 \Omega_p&=& \frac{\sqrt{M}}{a\sqrt{M}+r_0^{3/2}}\,.\label{Omegap}
\end{eqnarray}
The four-velocity of the particle on a timelike geodesic reads $r_0^2m_pU^\alpha=((r_0^2+a^2)Q/\Delta+a(L_p-aE_p),0,0,L_p-aE_p+aQ/\Delta)$, where $\Delta=r^2-2Mr+a^2$, $Q=(r_0^2+a^2)E_p-aL_p$.

\noindent{\bf{\em IIB. Wave emission.}}
Because of the coupling to matter, the orbiting object emits both gravitational and scalar radiation. Gravitational radiation can be computed using Teukolsky's formalism \cite{Teukolsky:1973ha}. The relevant equations and their solution are presented by Detweiler \cite{Detweiler:1978ge}.
Here we focus on scalar wave emission. Defining
\be
\varphi(t,r,\Omega_p)=\sum_{l,m}\int d\omega e^{im\phi-i\omega t}\frac{X_{lm}(\omega,r)}{\sqrt{r^2+a^2}} S_{lm}(\theta)\,,
\ee
we get the non-homogeneous equation for the scalar field
\be
\left[\frac{d^2}{dr_*^2}+V\right]X_{lm\omega}(r)=\frac{\Delta}{(r^2+a^2)^{3/2}}T_{lm\omega}\,,\label{nonhom0}
\ee
where $dr/dr_*=\Delta/(r^2+a^2)$,
\be
T_{lm\omega}=-\frac{\alpha}{U^t}S^*_{lm}(\pi/2)\delta(r-r_0)\,m_p\delta(m\Omega_p-\omega)\label{tlmw}\,,
\ee
and the effective potential for wave propagation $V$ is given (e.g.) in \cite{Ohashi:1996uz}.
%
Let us consider two independent solutions $X_{lm\omega}^{r_+}$ and $X_{lm\omega}^{\infty}$ to the homogeneous equation satisfying the following boundary conditions:
\be
X_{lm\omega}^{\infty,r_+}\sim e^{ik_{{\infty},H} r_*} \quad
{\rm as}
\quad
r\to \infty,\,r_+\,,\nonumber
\ee
where $k_H=\omega-m\Omega_H$ and $k_{\infty}=\sqrt{\omega^2-\mu_s^2}$. Let
$W$ be their Wronskian.
The fluxes of scalar energy at the horizon and at infinity are
\beq
\dot E^{s}_{r_+,\infty}&=&m\Omega_p k_{H,\infty}|Z^{r_+,\infty}_{lm\omega}|^2\,,\label{scalar_fluxes}\\
Z_{lm\omega}^{r_+,\infty}&\equiv&-\alpha\frac{X_{lm\omega}^{\infty,r_+}(r_0)}{W U^t}\frac{S_{lm}^{*}(\pi/2)}{\sqrt{r_0^2+a^2}}\,m_p/M\,.
\eeq
%

\noindent{\bf{\em IIC. Analytic solution at low frequencies.}}
The scalar flux at infinity can be computed in the low-frequency regime \cite{Poisson:1993vp}. 
For $r_0/M\gg 1$ and $l=m=1$,
\be
\dot E^{s}_{\infty}=\frac{\alpha^2M^2}{12\pi}\frac{\left(1-\mu_s^2r_0^3/M\right)^{3/2}}{r_0^4}m_p^2
\Theta(\Omega_p-\mu_s)\,,\label{dipole}
\ee
where $\Theta(x)$ is the Heaviside function.
For generic modes, at large distances and for $\omega=m\Omega_p>\mu_s$, scalar radiation dominates over gravitational radiation: compare Eq.~(\ref{dipole}) with the standard quadrupole formula
$\dot E^{g}_\infty={32}/{5}\left({r_0}/{M}\right)^{-5}m_p^2/M^2$.
This result is oblivious to the presence of the rotating BH.
In fact, for $\omega>\mu_s$, the fluxes at the horizon are negligible. However, for frequencies close to $\mu_s$, a resonance occurs at~\cite{Detweiler:1980uk}:
\be
\omega_{\rm res}^2=\mu_s^2-\mu_s^2\left(\frac{\mu_s M}{l+1+n}\right)^2\,,\quad n=0,1,...\label{om_resonance}
\ee
From Eq.~\eqref{scalar_fluxes} we see that $\dot E^{s}_{r_+}<0$ in the superradiant regime ($k_H<0$). Close to resonance we get (cf. also~\cite{Detweiler:1980uk})
\be
X^{\infty}_{lm\omega}\sim r^{l+1}e^{-\mu_s^2Mr/(l+1+n)}\,.\label{behaviorresonance}
\ee
We have verified this result numerically, finding very good agreement with the analytical prediction.
For the fundamental mode $n=0$, at resonance, we find:
\be
W\sim i\sqrt{r_+^2+a^2}(r_+-r_-)^{l+iP}\frac{\Gamma[l+1] \Gamma[l+1-2 i P]}{\Gamma[2 l+1] \Gamma[1-2 i P]}\,,\nonumber
\ee
where $P=-2Mr_+k_H/(r_+-r_-)$.
Finally we can estimate the peak flux close to the resonant frequencies.
At large distances and for $l=m=1, \,n=0$ we find
\be
\dot{E}^{s, {\rm peak}}_{r_+}\sim -\frac{3 \alpha ^2  \sqrt{\f{r_0}{M}}m_p^2 M}{16\pi r_+\left(M^2-a^2\right)\left(\frac{a}{2r_+}-(\frac{M}{r_0})^{3/2}\right){\cal F}}
\,,\label{flux_ana_fin2}
\ee
%
with ${\cal F}=1+4 P^2$.
Quite surprisingly the scalar flux at the horizon {\it grows} in magnitude with $r_0$ and it is negative, due to superradiance, at sufficiently large distances (for generic $l$, the peak flux would scale as $\dot{E}^{s, {\rm peak}}_{r_+}\propto r_0^{2l-3/2})$. For very small $a$ the peak flux at resonance is instead positive, and it can also be very large: for the Schwarzschild geometry, $32 \pi M^4 \dot{E}^{s, {\rm peak}}_{r_+}\sim 
3\alpha ^2r_0^2m_p^2$.

\vspace{-0.5cm}
\begin{center}
\begin{table}
\begin{tabular}{cccc}
\hline
\hline
$\mu_s M$ & $\,\,\,r_0/M$ (resonance) & $\,\,\,(\alpha m_p/M)^{-2}\dot E^{s,\,{\rm peak}}_{r_+}$ & $\alpha_\text{crit}$\\
\hline 
$10^{-1}$	&	$4.33400288873563$	  &	$-0.1828$	& $1.1\cdot 10^{-1}$                     \\
$10^{-2}$	&	$21.4020987080510$	&	$-0.4881$	& $1.6\cdot 10^{-3}$                       \\
$10^{-3}$	&	$99.9339974413005$	&	$-1.1588$	& $2.3\cdot 10^{-5}$                       \\
\hline
\hline
\end{tabular}
\caption{\label{tab:resonance} 
Orbital radius at resonance and peak scalar flux for $n=0$, $l=m=1$, $a=0.99M$ and several values of $\mu_s M\ll1$. For comparison, a typical extreme mass ratio inspiral becomes detectable by space-based interferometers at radii $r_0/M\sim 50\,[(10^6M_\odot/M)(f_{\rm cut}/10^{-4}~{\rm Hz})]^{-2/3}$, where $f_{\rm cut}$ is the lower cutoff for the sensitivity threshold of the interferometer. A floating orbit occurs for $\alpha>\alpha_\text{crit}$. Notice that $\alpha_\text{crit}$ is well below current observational bounds~\cite{Perivolaropoulos:2009ak} for any $\mu_s$.}
\end{table}
\end{center}

\begin{figure}[ht]
\begin{center}
\centerline{\includegraphics[width=6.8cm,clip=true]{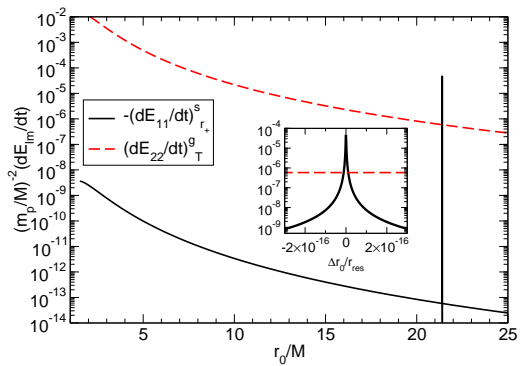}}
\caption{Dominant fluxes of scalar and gravitational energy ($l=m=1$ and $l=m=2$, respectively) for $\mu_s M=10^{-2}$, $\alpha=10^{-2}$ and $a=0.99M$. The inset is a zoom around resonance.
\label{fig:resonance}}
\end{center}
\end{figure}

\noindent{\bf{\em III. Floating orbits.}}
From the previous discussion it follows that, for any $\mu_s M\ll1$, there exists a frequency $\omega_{\rm res}\lesssim\mu_s$ for which the total flux $\dot E^{s}_\infty+\dot E^{s}_{r_+}+\dot E^{g}_\infty+\dot E^{g}_{r_+}=0$, because the negative scalar flux at the horizon is (in modulus) large enough to compensate for the other positive contributions. 
This expectation is confirmed by a full numerical integration of Teukolsky's equation: see Fig.~\ref{fig:resonance} and Table~\ref{tab:resonance}.
The width of the peak is proportional to the imaginary part of the resonant mode $\omega_I\propto\mu_s^{4l+5}$ \cite{Detweiler:1980uk}. For $l=1$, more explicitly,
\be
\omega_I=\mu_s\frac{(\mu_s M)^8}{24}\left(a/M-2\mu_s r_+\right)\,.\label{wi}
\ee
As $\mu_s\to0$ the imaginary part becomes tiny, and an accurate fine-tuning is needed to numerically resolve the resonance. For example, to resolve the peak at $r_0\sim100M$, corresponding to $\mu_s M=10^{-3}$, we tuned the location of $r_0$ to 25 decimal places. Computing the imaginary part of the unstable modes when $\mu_s\to0$ is also challenging, but we were able to obtain stable results for the resonance location and for the height of the peak. A fit to numerical results for $10M\lesssim r_0\lesssim100M$ (cf. Table \ref{tab:resonance}) yields $\dot E^{s,\,{\rm peak}}_{r_+}\sim r_0^{0.51}$, to be compared with $\dot E^{s,\,{\rm peak}}_{r_+}\sim r_0^{1/2}$ in  Eq.~\eqref{flux_ana_fin2}. Close to floating orbits
\be
\frac{d E_p}{dt}\sim-(E_p-E_f)\left.\frac{d\dot E_T}{dE_p}\right|_{E_p=E_f}\,,
\ee
where $E_f$ is the energy of the particle at the floating orbit, and we used the balance condition~\eqref{balance}. During inspiral, right before reaching the floating orbit, the time needed for the particle to increase its binding energy from $|E_f|-\epsilon$ to $|E_f|$ diverges logarithmically. Therefore, floating orbits are expected to last much longer than a typical inspiral timescale, with a potentially striking observational signature in the gravitational-wave spectrum. 

In the adiabatic approximation, the mass and angular momentum of the background spacetime are constant. However, the negative energy flux at the horizon reduces the BH mass and angular momentum ($\delta M<0$, $\delta J<0$). In order to estimate how long a particle can stay in a floating orbit we must go beyond the adiabatic approximation. Under ideal conditions, floating would stop only when the peak of the scalar flux at the horizon is too small to compensate for the gravitational flux,  $|\dot E^{g}|>|\dot E_\text{peak}^{s}|$.
From the balance condition~\eqref{balance} we find that $\delta E_p=0$, which, using Eq.~\eqref{Ep}, can be written as
\be
\delta r_0=\frac{\delta M r_0 \left(a^2+3(2 M-r_0) r_0+2 a \sqrt{r_0/M} (r_0-3M)\right)}{M \left(3 a^2-8 a \sqrt{M r_0}+(6 M-r_0) r_0\right)}\,,\nonumber
\ee
where we used the relation $\delta M=\Omega_p \delta J$, valid for circular orbits.
Substituting the equation above into Eq.~\eqref{flux_ana_fin2} and
approximating $\delta M/\delta t\sim \dot E^{s}_{r_+}=-\dot E^{g}_\infty=-32/5 \left(r_0/M\right)^{-5}m_p^2/M^2$ at resonance, we obtain, for $l=m=1$ and in the limits $\mu_s\to 0$ and $a\gg 1/\mu_s$,
\be
\frac{\delta\dot E^{s,\,{\rm peak}}_{r_+}}{\delta t}=-\frac{12\alpha^2}{5\pi}\frac{M}{a^2}\left(\f{m_p}{M}\right)^4\left(\frac{M}{r_0}\right)^{3}\,,\nonumber
\ee
which is negative: BH mass loss decreases the height of the peak on a timescale
\be
\frac{\dot E^{s,\,{\rm peak}}_{r_+}}
{\delta\dot E^{s,\,{\rm peak}}_{r_+}/\delta t}
\sim
\frac{5a}{32} \left(\frac{M}{m_p}\right)^2\left(\frac{r_0}{M}\right)^{7/2}\,,\nonumber
\ee
which does not depend on the coupling constant $\alpha$.

The delayed inspiral may have observational consequences. In particular, notice that in the absence of scalar fields the evolution of orbital frequency scales as 
\be
\dot\Omega_{\rm GR}/\Omega_p^2\sim 96/5(m_p/M)(M\Omega_p)^{5/3}\,.\label{inspiralGR}
\ee
Close to a floating orbit we find instead that 
\be
\dot\Omega_{\rm floating}/\Omega_p^2\sim 32(m_p/M)^2(M\Omega_p)^{7/3}\,.\label{inspiral_floating}
\ee
%
\noindent{\bf{\em IIIB. Perturbations of floating orbits.}}
It might be anticipated that floating orbits could be very sensitive to small perturbations, because the resonance has a small width proportional to $\omega_I$, which is given by Eq.~(\ref{wi}). One important class of perturbations consists of adding a small eccentricity $\varepsilon$ to the motion. As pointed out in \cite{Apostolatos:1993nu}, eccentricity will produce frequency sidebands. Because these sidebands are typically far away from resonance, scalar emission in these channels can be neglected, and we find that $\varepsilon$ tends to decrease close to floating orbits:
\be
\f{\dot{\varepsilon}}{\varepsilon}=
-\frac{352 m_p}{15 M^2}\left(\f{M}{r_0}\right)^4
\left(1-\f{991}{308}\f{M}{r_0}+{\cal O}(r_0^{-2})\right)\,,
\ee 
i.e. circular floating orbits stay circular. The argument breaks down at small separations, where a better understanding of eccentricity evolution would be desirable.

Perhaps the most dangerous perturbation would come from small ``kicks'' taking the object from a trajectory with frequency $\Omega_p$ to $\Omega_p +\delta \Omega_p$.
If $\delta\Omega_p \gg \omega_I$ (this is a conservative estimate, as floating itself extends over a larger portion of the frequency window) this might stop floating and superradiant energy extraction. When translated into perturbations of the position of the orbiting body, which we take to be of size $R_p$, floating would stop for
$\delta r_0/R_p \gg 10^{-5} \frac{M}{10^6 R_p}\left(\frac{r_0}{10M}\right)^{-11}$. For typical extreme-mass ratio inspirals of interest for space-based detectors $M/R_p \sim 10^6$ and $r_0\sim 10M$, so $\delta r_0/R_p\gg 10^{-5}$.

This rough estimate suggests that when $r_0/M\gg 1$, floating is easily stopped by small external perturbations. At large distance, even thermal fluctuations of the orbiting body may disrupt a floating orbit.
Note however that this estimate is overly pessimistic, as it considers only the fundamental mode. For the first few overtone numbers $n$ the scalar flux at peak is roughly independent of $n$. If this behavior holds at large $n$, then the effective allowed window $\delta \Omega_p$ would be larger by powers of $1/\mu_s^2$. Further investigation of this issue is necessary to understand the relevance of floating orbits.

The small orbiting body was assumed throughout to be point-like. In reality there will be tidal interactions with the field of the massive BH, which will alter the gravitational waveform. For extreme mass-ratios such as the one we consider, this effect is negligible \cite{Baumgarte:2004xq}. Also, tidal effects do not in any way affect the {\it existence} of floating orbits, but they may have an impact in the disruption of floating. It is also conceivable that, on such large timescales, nonlinear mode coupling could channel energy to higher-order modes, slowly washing out the peak in the scalar flux. An analysis of tidal effects and of the nonlinear regime is beyond the scope of this paper.

\noindent{\bf{\em IV. Sinking orbits.}}
For small rotation, long-lived modes of frequency $\omega\simeq \mu_s$ can be excited outside the innermost stable circular orbit (ISCO) if $\Omega_H<\omega<\Omega_\text{ISCO}$.
This can happen for $0\leq a\lesssim 0.36 M$, and the corresponding resonances are now {\it stable} (cf. the sign change in Eqs.~(\ref{flux_ana_fin2}) and (\ref{wi}) for $\mu_s>\Omega_H$). For these resonances, the scalar flux at the horizon is large and positive. For $r_0\gg M$ (or when $\mu_s M\ll1$) the scalar flux dominates over the gravitational flux. From Eq.~\eqref{flux_ana_fin2}, using $\delta E_p/\delta t=-\dot E_T\sim-{\dot E}^{s, {\rm peak}}_{r_+}$ and neglecting mass and angular momentum loss, we get (for $l=m=1$, $n=0$)
\begin{equation}
\frac{\dot\Omega_{\rm sinking}}{\Omega_p^2}=\frac{9\alpha^2 m_p/M}{8\pi (1-a^2/M^2)(2 r_+\Omega_p-a/M){\cal F}}(M\Omega_p)^{-2}\,.\nn
\end{equation}
These resonances are not superradiant in origin. In fact, they are present even for nonrotating BHs. For very small $a$, 
$\dot\Omega_{\rm sinking}/\Omega_p^2\sim 18\alpha^2 m_p(M\Omega_p)^{-3}/(256 M)$ close to resonance: compare the inspiral rate in general relativity without scalar fields, Eq.~(\ref{inspiralGR}). Thus, close to stable resonances the orbiting body inspirals much faster, and the orbit {\it sinks}.
Although this effect seems huge at small frequencies, its timescale is extremely small, since the resonance width $\sim\omega_I$ (cf. Eq.~\eqref{wi}). Indeed, we get
\begin{equation}
\tau_s\sim\left|\frac{\omega_I}{{\dot\Omega_p}}\right|=\frac{\pi}{27}\frac{(M^2-a^2){\cal F}}{m_p}\left(2r_+\mu_s-\frac{a}{M}\right)^2\frac{(M\mu_s)^9}{\alpha^2}\,.\nonumber
\end{equation}
Note that $\tau_s=0$ for a normal mode, at $\Omega_p\sim\mu_s=\Omega_H$.

\noindent{\bf{\em V. Conclusions.}}
We described an extreme form of energy extraction from a Kerr BH when a massive scalar field is coupled to a point particle in circular orbit around the BH. This is the first example of a phenomenon produced by a resonance between orbital frequencies and proper oscillation frequencies of the BH. Our results apply in principle to stationary background geometries different from the Kerr solution. The existence of an ergoregion is mandatory. 
There are few experimental bounds on $\mu_s$. Because of superradiance, Kerr BHs
are unstable under massive scalar perturbations, so (in principle) the
observation of fast-spinning astrophysical BHs can exclude the existence of
very light scalars (axions) \cite{Arvanitaki:2010sy}.
The strongest instability was found to occur when
$\mu_s=5.63\times 10^{-11}\left(M/M_\odot\right)^{-1}$~eV on a
timescale $\tau = 32.8\,(M/M_\odot)\,{\rm s}$ \cite{Cardoso:2005vk}. 
Our results apply to generic theories of gravity, a particular instance being
BD theory. Floating may occur and be observed at radii $r_0<50M$
(cf. caption of Table I). From Eq.~\eqref{flux_ana_fin2}, imposing $\dot{E}^{s, {\rm peak}}>\dot E^g$ at $r_0=50M$, we find the smallest value of $\alpha$ which allows for observable effects. This translates into $\omega_{\rm BD}\lesssim
10^8$~$(2\cdot 10^8)$ for $a/M=0.99$~($a/M=0.5$). 
Thus, floating orbits could in principle provide much more stringent constraints on BD theory than those coming from Solar System observations.


Current searches for gravitational waves from coalescing compact binaries use post-Newtonian waveform templates, and are strongly biased towards general relativity. If light scalar degrees of freedom couple to matter, binaries may merge in a much more interesting way, and current searches based on matched-filtering techniques may underperform. In particular, gravitational waveforms would carry a clear signature of floating orbits: compare Eq.~(\ref{inspiral_floating}) with the standard general relativistic prediction of Eq.~(\ref{inspiralGR}). Future space-based and advanced Earth-based detectors may have the potential to reveal the existence of floating orbits and of other surprising ``anomalies'' in the orbital evolution of compact binaries.

\vspace{0.1cm}
\noindent{\em Acknowledgments.}
We thank Ana Sousa for preparing Fig.~1. We thank Universidade Federal do Par\'a, ICMS and the Pedro Pascual Center of Science for hospitality while parts of this work were completed. We thank J.~Alsing, E.~Barausse, C.~M.~Will and all the participants of the workshops ``ExDiP2010 -- Extra-Dimension Probe by Cosmophysics'', ``Numerical relativity beyond astrophysics'' and ``Gravity -- New perspectives from strings and higher dimensions'' for useful comments and suggestions. This work was supported by the {\it DyBHo--256667} ERC Starting Grant, by NSF Grant PHY-0900735, by NSF CAREER Grant PHY-1055103, and by FCT - Portugal through PTDC projects FIS/098025/2008, FIS/098032/2008, CTE-AST/098034/2008.



\begin{thebibliography}{99}

\bibitem{Arvanitaki:2009fg}
  A.~Arvanitaki, S.~Dimopoulos, S.~Dubovsky, N.~Kaloper, J.~March-Russell,
  Phys.\ Rev.\  {\bf D81}, 123530 (2010).

\bibitem{Arvanitaki:2010sy}
  A.~Arvanitaki, S.~Dubovsky,
  Phys.\ Rev.\  {\bf D83}, 044026 (2011).

\bibitem{Kodama:2011zc}
  H.~Kodama and H.~Yoshino,
  arXiv:1108.1365 [hep-th].

\bibitem{Wagoner:1970vr}
  R.~V.~Wagoner,
  Phys.\ Rev.\  {\bf D1}, 3209-3216 (1970).

\bibitem{Hersh:1985hz}
  J.~Hersh, R.~Ove,
  Phys.\ Lett.\  {\bf B156}, 305 (1985).

\bibitem{DeFelice:2010aj}
  A.~De Felice and S.~Tsujikawa,
  Living Rev.\ Rel.\  {\bf 13}, 3 (2010).

\bibitem{zeldovich} Ya. B. Zel'dovich,
Pis'ma Zh. Eksp. Teor. Fiz. {\bf 14}, 270 (1971) [JETP Lett. {\bf
14}, 180 (1971)]; Zh. Eksp. Teor. Fiz {\bf 62}, 2076 (1972) [Sov.
Phys. JETP {\bf 35}, 1085 (1972)].

\bibitem{Bekenstein:1973mi}
  J.~D.~Bekenstein,
  Phys.\ Rev.\  {\bf D7}, 949-953 (1973).
  
\bibitem{Misner:1972kx}
  C.~W.~Misner,
  Phys.\ Rev.\ Lett.\  {\bf 28}, 994-997 (1972).
  
\bibitem{Press:1972zz}
  W.~H.~Press, S.~A.~Teukolsky,
  Nature {\bf 238}, 211-212 (1972).
  
\bibitem{Damour:1976kh}
  T.~Damour, N.~Deruelle, R.~Ruffini,
  Lett.\ Nuovo Cim.\  {\bf 15}, 257-262 (1976).
  
\bibitem{Cardoso:2004nk}
   V.~Cardoso, O.~J.~C.~Dias, J.~P.~S.~Lemos, S.~Yoshida,
  Phys.\ Rev.\  {\bf D70}, 044039 (2004).

\bibitem{Cardoso:2005vk} V.~Cardoso, S.~Yoshida,
  JHEP {\bf 0507}, 009 (2005); S.~R.~Dolan,
Phys.\ Rev.\  {\bf D76}, 084001 (2007);
J.~G.~Rosa,  
JHEP {\bf 1006}, 015 (2010);
S. Hod, 
Phys.\ Rev.\ {\bf D84}, 044046 (2011).

\bibitem{Cardoso:2004hs}
  V.~Cardoso, O.~J.~C.~Dias,
  Phys.\ Rev.\  {\bf D70}, 084011 (2004); 
  H.~Kodama, R.~A.~Konoplya, A.~Zhidenko,
  Phys.\ Rev.\  {\bf D79}, 044003 (2009).

\bibitem{Friedman:1978} J.L. Friedman, 
Commun. Math. Phys. {\bf 63}, 243 (1978).

\bibitem{Cardoso:2007az}
  V.~Cardoso, P.~Pani, M.~Cadoni, M.~Cavaglia,
  Phys.\ Rev.\  {\bf D77}, 124044 (2008); V.~Cardoso, P.~Pani, M.~Cadoni, M.~Cavaglia, 
  Class.\ Quant.\ Grav.\  {\bf 25}, 195010 (2008);  P.~Pani, E.~Barausse, E.~Berti, V.~Cardoso,
  Phys.\ Rev.\  {\bf D82}, 044009 (2010);
  V.~Cardoso, O.~J.~C.~Dias, J.~L.~Hovdebo, R.~C.~Myers,
  Phys.\ Rev.\  {\bf D73}, 064031 (2006).

\bibitem{Thorne:1971}
  K.~S.~Thorne, J.~J.~Dykla, 
  Astrophys.\ J.\  {\bf 166}, L35-L38 (1971).

\bibitem{Psaltis:2007cw}
  D.~Psaltis, D.~Perrodin, K.~R.~Dienes, I.~Mocioiu,
  Phys.\ Rev.\ Lett.\  {\bf 100}, 091101 (2008).
    
\bibitem{Ohashi:1996uz}
  A.~Ohashi, H.~Tagoshi, M.~Sasaki,
  Prog.\ Theor.\ Phys.\  {\bf 96}, 713-727 (1996).
  
\bibitem{Will:2005va}
  C.~M.~Will,
  Living Rev.\ Rel.\  {\bf 9}, 3 (2006).

\bibitem{Perivolaropoulos:2009ak}
  L.~Perivolaropoulos,
  Phys.\ Rev.\  {\bf D81}, 047501 (2010).
  P.~J.~Steinhardt, C.~M.~Will,
  Phys.\ Rev.\  {\bf D52}, 628-639 (1995).

\bibitem{Teukolsky:1973ha}
  S.~A.~Teukolsky,
  Astrophys.\ J.\  {\bf 185}, 635-647 (1973).
  
\bibitem{Detweiler:1978ge}
  S.~L.~Detweiler,
  Astrophys.\ J.\  {\bf 225}, 687-693 (1978).
  
\bibitem{Poisson:1993vp}
  E.~Poisson,
  Phys.\ Rev.\  {\bf D47}, 1497-1510 (1993).
  
\bibitem{Detweiler:1980uk}
  S.~L.~Detweiler,
  Phys.\ Rev.\  {\bf D22}, 2323-2326 (1980).  

\bibitem{Apostolatos:1993nu}
  T.~Apostolatos, D.~Kennefick, E.~Poisson, A.~Ori,
  Phys.\ Rev.\  {\bf D47}, 5376-5388 (1993).

\bibitem{Baumgarte:2004xq}
  T.~W.~Baumgarte, M.~L.~Skoge, S.~L.~Shapiro,
  Phys.\ Rev.\  {\bf D70}, 064040 (2004).
  [gr-qc/0405077].
  

\end{thebibliography}
\end{document}